# Aligning Models with Their Realization through Model-based Systems Engineering

Lovis Justin Immanuel Zenz, Erik Heiland, Peter Hillmann, and Andreas Karcher

Institute for Applied Computer Science
University of the Bundeswehr Munich
Neubiberg, Germany
{lovis.zenz; erik.heiland; peter.hillmann; andreas.karcher}@unibw.de

*Abstract*—In this paper, we propose a method for aligning models with their realization through the application of model-based systems engineering. Our approach is divided into three steps. (1) Firstly, we leverage domain expertise and the Unified Architecture Framework to establish a reference model that fundamentally describes some domain. (2) Subsequently, we instantiate the reference model as specific models tailored to different scenarios within the domain. (3) Finally, we incorporate corresponding run logic directly into both the reference model and the specific models. In total, we thus provide a practical means to ensure that every implementation result is justified by business demand. We demonstrate our approach using the example of maritime object detection as a specific application (specific model / implementation element) of automatic target recognition as a service reoccurring in various forms (reference model element). Our approach facilitates a more seamless integration of models and implementation, fostering enhanced Business-IT alignment.

## I. MOTIVATION

Models constitute an important part of enterprise architecture (EA) management. Throughout an EA project, it is important to generate diagrams that correspond to each other. Only in this way can the model composed of them be consistent. Furthermore, when there is a reference model representing some service reoccurring multiple times throughout an organization, the specific models representing instances of this service – and by extend their implementations – need to correspond to the reference model [1][2]. This ensures that a coherent overall picture emerges. Thus, the implicitly expected reusability of reference model contents is given.

Hence, the question arises how the modelling process can be supported in such a way that the desired correspondence is achieved. In this paper, we propose such an approach that employs model-based systems engineering (MBSE) [3][4] to align models with their implementation. Fig. 1 visualizes an overview of the approach with deliverables lined up in the middle. Primary inputs are placed above the deliverables while a secondary input is positioned below them. Finally, the overarching MBSE-based process is portrayed with block arrows.

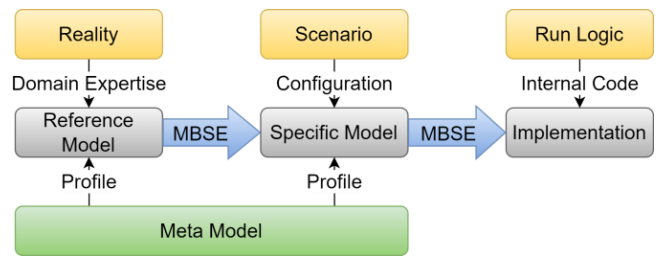

Fig. 1. Overview of Approach

Regarding modelling, we distinguish between the following concepts:

1) The term 'model' denotes the comprehensive structure that is created.
2) This structure organizes one or more views, which display model elements and relations between them in some arrangement, within a package/folder structure. We refer to such a view as 'diagram'.
3) When a diagram's arrangement is prescribed by a viewpoint originating from a framework, we use the notation "*<name>*' diagram", where *<name>* is replaced with the name of the respective viewpoint.

The remainder of our paper is structured in the following way. In Section II, we analyze the baseline scenario of our project to identify requirements for our approach. In Section III, we next summarize existing work related to our approach and its goals. In Section IV, we then present our concept. In Section V, we subsequently demonstrate our approach based on a real-world example in maritime context. In Section VI, we offer our conclusion.

## II. REQUIREMENTS ANALYSIS

Before an approach for aligning models with their implementation can be developed, relevant quality criteria need to be collected in the form of requirements [5]. We formalize these requirements based on the framework criteria of our project as derived from the following scenario.

In the field of automatic target recognition (ATR), various services exist. Due to their diverse objectives, each of them is developed as an isolated solution. This diversity is reflected in the utilization of various techniques such as sonar, lidar and image-based machine learning. Despite the differences in techniques, the fundamental steps – preprocessing, detection, classification, etc. – required to realize an ATR service remain highly similar. Therefore, the development of a reference model for ATR would offer significant benefits. Aligning specific models, along with their implementations, to this reference model would facilitate the identification of commonalities among instances. Consequently, the knowledge gained from developing one instance could be effortlessly leveraged to benefit other instances.

The following requirements regarding the reference model emerge:

- The reference model shall conform to the *Unified Architecture Framework (UAF)*.
- The reference model shall reflect domain expertise.

Furthermore, the following requirements regarding the tool chain emerge:

- The tool chain shall be uninterrupted. Ideally, the entire tool chain shall be realized within one tool. If this is not feasible, all transitions between tools shall be seamless enough that the user retains the impression of a continuous tool chain.
- The tool chain in its entirety shall be capable of handling demanding implementations. For instance, solving complex mathematical problems shall be possible.
- The tool chain shall allow for implementations including external code in addition to internal code. Both locally and remotely hosted external code shall be supported.
- The tool chain shall be able to handle implementations that incorporate both local and remote external data. This shall hold true with respect to local files as well as URL endpoints and databases.
- The tool chain shall be able to facilitate dynamic implementations by supporting user input.

## III. RELATED WORK

There are several previous works that impact the approach presented in this paper. We examine these publications in terms of what information is obtained from each and how our work differs from theirs, respectively.

Sparx Systems provides extensive instructions for utilizing *Enterprise Architect (Sparx EA)* to model diagrams and integrate code into diagram elements. Furthermore, they provide an introduction to generating, building and running the integrated code to achieve simulations of model-based implementations [6] - [16]. From this information, we extract a pool of techniques as well as basic insights into their implementation. Upon examining the *Sparx EA User Guide Series*, it becomes apparent that several of the therein proposed procedures require adaptation or may not function at all when *Sparx EA 16.0* is employed. A more in-depth consideration of the pool of techniques follows in Section IV.

The Object Management Group (OMG) furnishes *UAF* as a means for supporting enterprise architecture development [17]. For our approach, *UAF* is especially interesting due to the following reasons:

- *UAF* is suitable for modelling systems of systems. To this end, it can be used in conjunction with the *Systems Modelling Language (SysML)* [18].
- Moreover, *UAF* enables holistic modelling from the abstract strategy level down to the detailed resource level [17].

From these norms, we extract the *UAF* grid [17], which is displayed in Fig. 2, as well as the intended contents of its cells. A frame highlights the parts we focus on.

| | Taxonomy Tx | Structure Sr | Connectivity Cn |
|---|---|---|---|
| Metadata[a] Md | Metadata Taxonomy Md-Tx[f] | Metadata Structure Md-Sr | Metadata Connectivity Md-Cn |
| Strategic St | Strategic Taxonomy St-Tx | Strategic Structure St-Sr | Strategic Connectivity St-Cn |
| Operational Op | Operational Taxonomy Op-Tx | Operational Structure Op-Sr | Operational Connectivity Op-Cn |
| Services Sv | Service Taxonomy Sv-Tx | Service Structure Sv-Sr | Service Connectivity Sv-Cn |
| Personnel Pr | Personnel Taxonomy Pr-Tx | Personnel Structure Pr-Sr | Personnel Connectivity Pr-Cn |
| Resources Rs | Resource Taxonomy Rs-Tx | Resource Structure Rs-Sr | Resource Connectivity Rs-Cn |

Fig. 2. Excerpt of the UAF 1.1 Grid

The Object Management Group additionally provides *Model Driven Architecture (MDA)* as an approach for developing software using models throughout the whole process [19]. Rhazali collects multiple approaches that extend *MDA* [20]. While these approaches provide several ways to generate running programs from models, regarding the models, they mainly focus on the resource level. Hence, no holistic perspective is provided. From these publications, we extract *MDA* as an approach to build upon.

Beery and Paulo address the application of MBSE to mission engineering [21]. When discussing the generation of linked models and simulations, they focus on the connection between operational and resource artifacts. Thereby, they do not consider the generation of operational model contents based on higher abstraction model contents. Furthermore, they do not adhere to a framework [22] for structuring their models.

Holt et al. utilize MBSE to accomplish requirements engineering [23]. Although they suggest a suitable approach for conducting model-based requirements engineering (MBRE) and thereby focus on higher levels of abstraction, they do not consider the entire spectrum of the challenge from a strategic perspective down to a resource perspective.

Sunkle, Kulkarni, and Roychoudhury emphasize the significance of structured decision making and holistic perspectives in EA management (EAM) [24]. They propose an approach based on a combination of common EA models and intentional models to achieve this. Although their approach offers detailed solutions, which allow for iterative improvement of existing EA models, it still requires EA models describing the initial state. Hence, the challenge of consistently creating those large-scale models remains. Sunkle and Rathod extend the above approach by exploring visual modelling support [25]. Thereby, they manage to simplify the modelling process. Nevertheless, the requirement for initial EA models and the challenge of consistently creating them remain. Sunkle et al. reinforce our belief that domain expertise is essential for generating useful EA models.

## IV. CONCEPT

Our approach conceptually originates from the baseline in Fig. 3. Considering the domain of some reoccurring service, we aspire comprehensive modelling from a strategic down to a resource level. Thereby, we obtain a strong foundation to derive a matching implementation afterwards. Instantiations of the reoccurring service shall reflect the requirements of specific scenarios and use cases. Moreover, each stakeholder shall be provided with diagrams corresponding to their respective perspective. To facilitate replaceability, especially on the resource level, interfaces shall be taken into account.

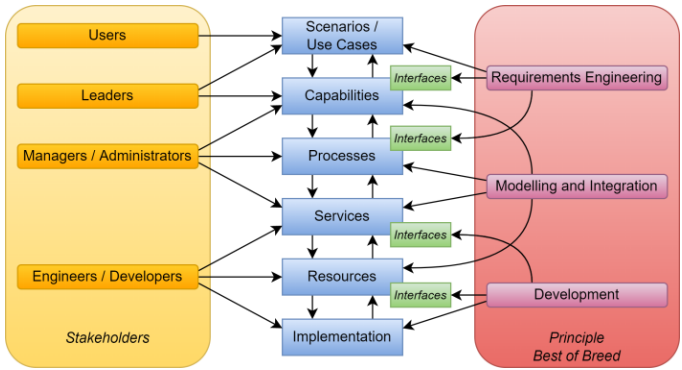

Fig. 3. Concept Baseline

Our analysis of related work provides a pool of techniques employable to realize the modelling. This pool is exhibited in Fig. 4. Based on the identified requirements, we select the techniques to actually employ. These techniques are highlighted in darker shades.

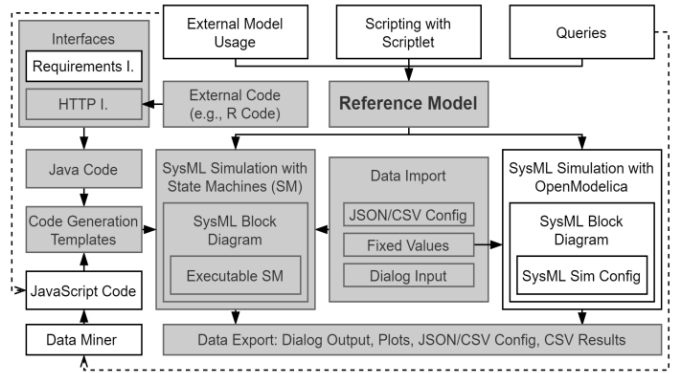

Fig. 4. Pool of Techniques

Furthermore, the identified requirements influence various design decisions regarding the specific utilization of the selected techniques. Altogether, this results in our concept with the following top-level steps:

1) We employ MBSE to create the reference model in accordance with *UAF*. Thereby, we limit ourselves to the *UAF* layers 'Strategic', 'Operational', 'Services', and 'Resources'.
2) Based on the resulting reference model, we employ MBSE again to derive a specific model from its 'Resources' layer. We add a diagram to the specific model that represents the configuration of the intended implementation and contains all internal data required for it.
3) We add an artifact to the diagram. This artifact holds the required meta information to generate code from the configuration diagram, build that code and run the result.

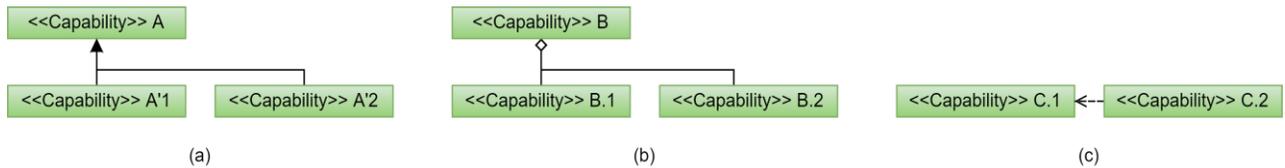

Fig. 5. Example Patterns for Strategic Taxonomy (a), Strategic Structure (b), and Strategic Connectivity (c)

The following steps are performed to create the reference model:
1) On its 'Strategic' layer, 'Capability' blocks are created for each desired capability. They are presented in 'Strategic Taxonomy', 'Strategic Structure' and 'Strategic Connectivity' diagrams – see Fig. 5.
2) On its 'Operational' layer, 'Operational Activity' blocks are derived from the 'Capability' blocks. 'Operational Performer' blocks are added here as well. The presentation yields results similar to the ones seen in Fig. 5.

3) On its 'Services' layer, 'Service Specification' blocks are derived from the 'Operational Activity' blocks. When applicable, they are divided into 'Service Functions'. Again, the presentation yields results similar to the ones seen in Fig. 5.
4) On its 'Resources' layer, 'System' blocks are derived from the 'Service Specification' and 'Service Function' blocks. If several 'System' blocks represent different resources that can be alternatively utilized to provide services represented by the same 'Service Specification' or 'Service Function' block, these 'System' blocks are structured in an inheritance tree. One more time, the presentation yields results similar to the ones seen in Fig. 5.

5) Finally, the underlying functionality of each non-abstract 'System' block is added to it in the form of internal code. Thereby, it is necessary to decide on the programming language used for describing the functionality of each block. The following steps are simplified if the same language is used throughout the whole process.

To ensure replaceability between 'System' blocks, which can be alternatively utilized as mentioned above, internal code must be designed to support any permissible combination of 'System' blocks. This can be achieved by specifying a design pattern for a group of 'System' blocks and adhering to this pattern when writing the internal code for each block within the group.

Furthermore, to allow for incorporating external functionality and data in the implementation to be generated out of the specific model, which is again generated out of this reference model, interfaces can be employed. First and foremost, *HTTP* interfaces accessing *RESTful API* endpoints are suitable to achieve this.

To derive the configuration diagram, 'System' blocks from the reference model are embedded as linked objects. Each embedded block represents a logically encapsulated component of the implementation – e.g., a class in *Java* terms. Whenever there is a set of blocks, each of which could be employed to achieve the same result in a different way, the appropriate block is chosen based on the intended implementation configuration.

When adding the artifact, each 'System' block within the configuration needs to be added to it as a property. Furthermore, the previously chosen programming language needs to be specified in the artifact's preferences. As the artifact's language preference and the actually chosen programming language have to match, several artifacts are required if the implementation shall consist of several parts employing different programming languages. This reflects the fact that each such part of the implementation needs to be generated, built, and run separately.

## V. RESULTING SPECIFIC MODEL AND IMPLEMENTATION

The approach presented in Section IV can be employed to achieve alignment between models and their implementation. In this section, we show this for an exemplary case, the associated specific model and the implementation generated from it.

### A. Exemplary Case

The following simulation scenario in the context of the general scenario from Section II constitutes the exemplary basis for our evaluation. For its specification, we employ the formula symbols in Table I. Therein and in the following, *AUV* abbreviates *Autonomous Underwater Vehicle*, *MCU* abbreviates *Movement Control Unit*, and *TCU* abbreviates *Target Classification Unit*. Among these symbols, $t$ is a parameter, whereas all symbols with ($t$) are variables. All other symbols are constants. The latter especially include $t_0$, $t_i$ and $t_n$. Furthermore, we assume

$$t \in T \subset \mathbb{N}_0, |T| < \infty \quad (1)$$

and

$$\forall v(t) \in \{v_{actual}(t), v_{active}(t)\}, t < t_n, u \in [0,1) : \\ v(t) = v(t+u). \quad (2)$$

An *AUV* is intended to move with $v_{desired}$ from $p_{desired}(t_0)$ to $p_{desired}(t_n)$. Hence, it starts moving at $p_{actual}(t_0)$ with

$$v_{active}(t_0) := v_{desired}. \quad (3)$$

We assume

$$p_{actual}(t_0) = p_{desired}(t_0). \quad (4)$$

TABLE I  FORMULA SYMBOLS USED IN THE EXEMPLARY CASE

| Parameter | Meaning |
|---|---|
| $t$ | Point in time within the simulation |
| $t_0$ | Starting time of the simulation / first $t$ |
| $t_i$ | Waiting time of the *MCU* until its activation |
| $t_n$ | Duration of the simulation / last $t$ |
| $T$ | Set of all $t$ |
| $\delta t$ | Time step / distance between two points in time |
| $p_{desired}(t)$ | Desired position of the *AUV* at $t$ |
| $p_{actual}(t)$ | Actual position of the *AUV* at $t$ |
| $p_{deviation}(t)$ | Position deviation of the *AUV* at $t$ |
| $v_{desired}$ | Desired velocity of the *AUV* |
| $v_{actual}(t)$ | Actual velocity of the *AUV* at $t$ |
| $v_{active}(t)$ | Active velocity of the *AUV* at $t$ |
| $v_{passive}$ | Passive velocity of the *AUV* |
| $h$ | Threshold of the *TCU* |
| $s_j$ | Strength of signal $j$ |
| $N(t)$ | Background noise at $t$ |
| $N_0$ | Background noise with inactive *MCU* |
| $\delta N$ | Background noise increase due to active *MCU* |

A current impacts the *AUV* such that it is additionally constantly moved with $v_{passive}$.

Hence, the total movement of the *AUV* comes down to

$$v_{actual}(t) := v_{active}(t) + v_{passive}, \forall t \in T. \quad (5)$$

It can be seen that

$$v_{passive} \neq (0,0,0) \implies v_{actual}(t_0) \neq v_{desired}. \quad (6)$$

Therefore, due to position change of the *AUV*, there is

$$p_{deviation}(t) := p_{actual}(t) - p_{desired}(t). \quad (7)$$

From

$$t_i \in [t_0, t_n] \quad (8)$$

on, a *MCU* is activated, which means that it starts correcting the motion of the *AUV* such that

$$p_{actual}(t_n) = p_{desired}(t_n) \quad (9)$$

by setting

$$v_{active}(t) := v_{desired} - v_{passive} - \frac{p_{deviation}(t)}{\delta t}, \quad (10)$$
$$\forall t \geq t_i.$$

Hence,
$$v_{active}(t) = v_{desired}, \forall t < t_i. \quad (11)$$

Furthermore, a *TCU* receives one signal
$$j \in [0, m] \quad (12)$$

with corresponding $s_j$, each from several targets. Each of these targets may either be a wanted object or not. The targets are classified in accordance with Table II.

TABLE II  DECISION MAKING OF THE TARGET CLASSIFICATION UNIT

| Condition | Decision |
|---|---|
| $s_j + N(t) \geq h$ | Target is a wanted object |
| $s_j + N(t) < h$ | Target is another object |

As soon as it is activated, the *MCU* negatively impacts the *TCU* by causing $\delta N$ such that

$$N(t) = \begin{cases} N_0 & \text{for } t < t_i \quad (13a) \\ N_0 + \delta N & \text{for } t \geq t_i. \quad (13b) \end{cases}$$

This can impact occurrence chances of the errors shown in Table III. More specifically, the chance of false positives might be increased while the chance of false negatives might be reduced.

TABLE III  POSSIBLE ERRORS OF THE TARGET CLASSIFICATION UNIT

| Error | Meaning |
|---|---|
| False positive | Target is wrongly classified as a desired object |
| False negative | Target is wrongly classified as a non-desired object |

Based on the occurring classification errors, the suitability of assignments of $t_i$ and $h$ for a particular scenario defined by the remaining constants and some of the variables is determined.

### B. Specific Model

By modelling with *Sparx EA*, we can build upon already existing work stored within a *Sparx EA* repository [26]. From a reference model previously generated in accordance with the procedure described in Section IV, we derive a specific model containing the configuration diagram depicted in simplified form in Fig. 6. The diagram represents the exemplary case detailed in Subsection V.A. To this end, it primarily includes transferred 'System' blocks, to which 'Software' blocks as well as general blocks, and an 'executable statemachine' artifact are added. Including the additional elements enables the integrated simulation illustrated in Subsection V.C.

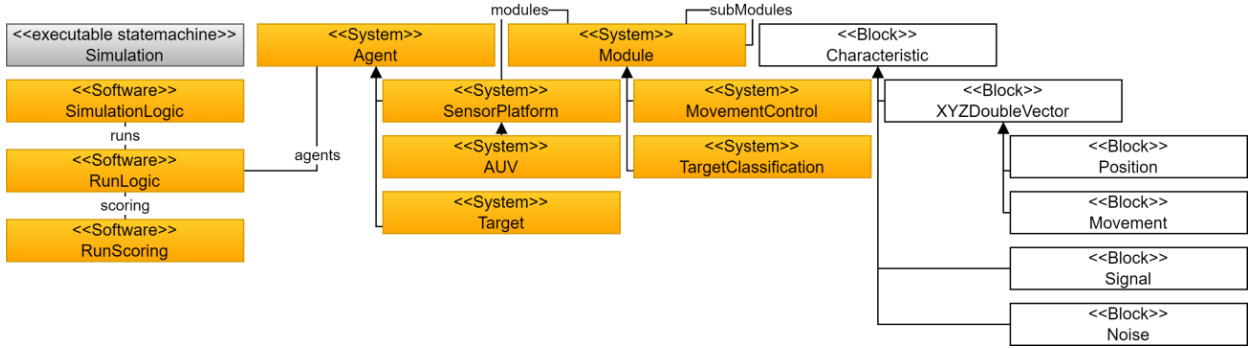

Fig. 6.  Simplified Configuration Diagram for Exemplary Case

As the configuration diagram is part of a specific model instead of the reference model, we employ the generic '*SysML* Block Definition' template for its creation.

### C. Typical Implementation Run

The simulation corresponding to the exemplary case presented above can be directly started from within the specific model. Assuming the utilization of *Sparx EA*, after opening the configuration diagram and highlighting the artifact, this can be triggered by selecting "*Simulate > Executable States > Statemachine > Execute > Generate, Build and Run*" from the header menu bar. Subsequently, following a short waiting time, during which the code is generated and built, the actual execution of the simulation begins. Firstly, variable simulation parameters can be entered by the user. Next, the user is informed about constant simulation parameters. An example for a consistent set of simulation parameters – both variable and constant ones – is listed in Table IV.

TABLE IV.  EXEMPLARY PARAMETER SET

| Reference Point | Parameter | Value |
|---|---|---|
| AUV | $t_n$ | 5 |
| | $p_{actual}(t_0)$ | (0,0,0) |
| | $v_{desired}(t_0)$ | (2,0,0) |
| | $v_{passive}$ | (0,1,0) |
| MCU | $t_i$ | 2 |
| | $\delta N$ | 1 |
| TCU | $h$ | 3 |
| | $N_0$ | 0 |
| Desired Target | $s_0$ | 3 |
| Undesired Target | $s_1$ | 2 |

Therein, vectors (*x*,*y*,*z*) with *x* denoting the width coordinate/change, *y* denoting the height coordinate/change and $z$ denoting the depth coordinate/change are employed to specify a position and two velocities. Then, the user is notified about the progress of the simulation in several windows. There is a window for each

$$t \in T. \tag{14}$$

Finally, the amounts of false positives and false negatives are reported to the user. The simulation terminates as soon as the user dismisses the scoring report. Figure 7 illustrates the corresponding scoring results. Notably, zero false negatives and three false positives occur in total. The first false positive occurring for $t = 3$ suggests that the false positives result from the activation of the *MCU*.

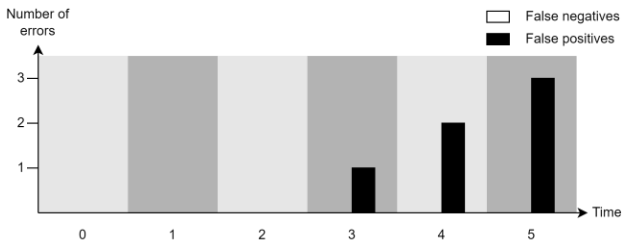

Fig. 7. Scoring Results for Exemplary Parameter Set

## VI. CONCLUSION

It can be seen that the specific model with its configuration diagram as shown in Subsection V.B matches the description of the exemplary case from Subsection V.A. This was achieved by employing MBSE while considering domain expertise. By directly integrating program logic into the specific model, we ensure that the implementation is consistent with the specific model and thus with the exemplary case. Because the specific model is derived from the reference model, the former complies with the latter. Therefore, the implementation, which is generated from the specific model, also complies with the reference model. Altogether, the goal of aligning models with their implementation has been reached.

The underlying concept offers a streamlined and holistic way to obtain this alignment. As the process begins with a reference model, already existing code can be reused for implementations reflecting different scenarios. Considering the above example of maritime object detection, future development of ATR services can leverage existing instances rather than beginning anew. Overall, our approach caters to requirements of modularity and reusability in the modelling and development process.

Nevertheless, the process still relies on a combination of modelling skills and domain expertise. Eliminating the need for domain expertise altogether may prove difficult. However, developing an IT solution to support or fully automate the modelling process appears promising. This would enable domain experts to model complex systems in accordance with appropriate meta models without requiring a deep understanding of the modelling process.